\documentclass[9pt]{elife}

\usepackage{lipsum} 
\usepackage{siunitx}
\usepackage{pdfpages}
\usepackage{float}
\usepackage{bbm}
\usepackage{amsmath}
\usepackage{adjustbox}
\usepackage{algorithm}
\usepackage[noend]{algpseudocode}
\usepackage{amsfonts}
\usepackage{tabularx}
\usepackage{booktabs}
\usepackage{pifont}
\usepackage{caption}
\usepackage{amsmath}
\usepackage{graphicx}

\DeclareSIUnit\Molar{M}

\title{Learning to combine top-down context and feed-forward representations under ambiguity with apical and basal dendrites}

\author[1,3*\authfn{1}]{Nizar Islah}
\author[1,2,3\authfn{1}]{Guillaume Etter}
\author[2,3\authfn{1}]{Mashbayar Tugsbayar}
\author[1,2,3]{Busra Tugce Gurbuz}
\author[2,3]{Blake Richards}
\author[1,3*]{Eilif B. Muller}
\affil[1]{Université de Montréal, CHU Ste-Justine Research Center, Montréal, Canada}
\affil[2]{McGill University, Montréal, Canada}
\affil[3]{Mila Quebec AI Institute, Montréal, Canada}

\corr{eilif.muller@umontreal}{EBM}
\corr{nizar.islah@umontreal.ca}{NI}

\contrib[\authfn{1}]{These authors contributed equally to this work}

\begin{document}

\maketitle

\begin{abstract}
One of the hallmark features of neocortical anatomy is the presence of extensive top-down projections into primary sensory areas, with many impinging on the distal apical dendrites of pyramidal neurons. While it is known that they exert a modulatory effect, altering the gain of responses, their functional role remains an active area of research.
It is hypothesized that one of the roles of these top-down projections is to carry contextual information that can help animals to resolve ambiguities in sensory data. 
One proposed mechanism of contextual integration is a combination of input streams at distinct (separate integration zone) apical and basal dendrites of pyramidal neurons.
Computationally, however, it is yet to be demonstrated how such an architecture could leverage distinct compartments for flexible contextual integration and sensory processing when both sensory and context signals can be unreliable.
Here, we implement an augmented deep neural network with distinct apical and basal compartments
that integrates a) contextual information from top-down projections to apical compartments, and b) sensory representations driven by bottom-up projections to basal compartments.
In addition, we develop a new multi-scenario contextual integration task using a generative image modeling approach. In addition to generalizing previous contextual integration tasks, it better captures the diversity of scenarios where neither contextual nor sensory information are fully reliable.
To solve this task, this model successfully learns to select among integration strategies. 
Specifically, when input stimuli and contextual information are contradictory, the performance of deep neural networks augmented with our "apical prior" exceeds that of single-compartment networks with otherwise equivalent architecture. We further show that our model can learn to integrate contextual information across time by using a task with input sequences and a gated recurrent unit layer to generate contextual signals.
Using layerwise relevance propagation, we extract the relevance of individual neurons to the model's predictions, revealing that a sparse subset of neurons encoding features of the context-relevant categories receive the largest magnitude of top-down signals when context and sensory input are aligned. We further show that this sparse gain modulation is necessary for best performance on the task.
Altogether, this suggests that the "apical prior" and the biophysically-inspired non-linear integration rule could be key components necessary for handling the ambiguities that animals encounter in the diverse contexts of the real world.
\end{abstract}

\section{Introduction}
Accurate perception relies on appropriate integration of context, as sensory signals contain incomplete or ambiguous information \citep{mumford_computational_1992, rao_predictive_1999}.
Contextual information can be derived from a number of sources, including the spatial and temporal domains \citep{eichenbaum_integration_2017}, task demands \citep{gilbert_top-down_2013}, as well as other sensory modalities.

Several key neuroanatomical and cellular features could support the computations associated with context integration.
In the mammalian brain, context has been proposed to be conveyed by top-down feedback pathways which are abundant in sensory regions of the neocortex \citep{felleman_distributed_1991, markov_anatomy_2014, harris_hierarchical_2019}.
Perturbations of top-down pathways from higher order areas have been associated with delayed object recognition \citep{kar_fast_2021} as well as disrupted stimulus response curves in lower-order regions \citep{wang_feedback_2007, nassi_corticocortical_2013}, which may reflect deficits in contextual processing.
Several feed-forward computational models of perception have implemented top-down pathways to enable object recognition of occluded images \citep{george_generative_2017, spoerer_recurrent_2017} as well as context-invariant perception \citep{naumann_invariant_2022}, and context-dependent gating for multi-task learning \citep{masse_alleviating_2018}.

In the neocortex, pyramidal neurons are thought to integrate feedforward and top-down, feedback information at their basal and apical dendrites, respectively \citep{larkum_dendritic_2007, spruston_pyramidal_2008}.
Notably, feedback has been proposed to modulate rather than drive neuronal activities \citep{sherman_actions_1998}.
This idea is supported by the unique morphological and physiological properties of pyramidal neurons, as activity in apical dendrites in vivo is generally not sufficient to drive somatic output responses alone \citep{stuart_determinants_1998, larkum_new_1999, larkum_cellular_2013, jarvis_neuronal_2018}.
However, activation of the apical compartment has been shown to act as a gain modulator, amplifying concurrent basal activity \citep{larkum_top-down_2004}, serving as a potential mechanism for contextual information.
While the physiological properties of pyramidal neurons have been described extensively (see \citet{spruston_pyramidal_2008} for review), there is no consensus on the exact mechanism by which top-down signals from higher-order regions modulate somatic firing rates and update sensory representations. 

Here, we propose a deep neural network architecture in which neurons are augmented specifically with apical and basal compartments and whose computation respects known biophysical properties of pyramidal neurons. We design a training paradigm such that the model updates its representations of sensory inputs arriving at the basal compartment
according to the top-down contextual information arriving at the apical compartment. 

In tandem with the proposed framework, we develop a multi-scenario task with a dataset containing ambiguous images that are mixtures of two image categories, in which ambiguity can only be correctly resolved by integrating contextual information with sensory representations.

We consider two versions of the problem setting. The first simplifies the contextual representation, which is given directly via a top-down oracle, and the model only needs to learn the apical parameters to correctly update the somatic activity. In the second, more challenging one, the model must use the temporal sequence as context and learn the appropriate contextual representations, along with the apical parameters to update somatic activity and ultimately solve the task. 

We show that the model learns to use apical modulation, driven by contextual inputs, to resolve sensory ambiguities.
To gain insights into the learned mechanisms of top-down modulation, we applied a standard method from the explainable AI literature (\citep{bach_pixel-wise_2015}) to identify the relative contributions of individual neurons and their apical compartments to network function.
We identified a subset of neurons that are highly relevant for the contextually-relevant category, and selectively amplified by top-down signals on apical dendrites.

Our results show that apical gain modulation of somatic activity by contextual information is a simple yet efficient way to solve new tasks, such as resolving perceptual ambiguity.  Moreover, the approach is flexible, because it is achieved without altering what the model has previously learned, as it does not introduce or modify any synapses in the feed-forward pathway.

These findings provide a candidate mechanism for how neocortical pyramidal neurons integrate top-down contextual information at their apical dendrites to support robust perception in the face of ambiguous data.

\section{Results}
 \subsection{Functional model of context integration in apical dendrites}
\begin{figure}[htp]
\includegraphics[width=\linewidth]{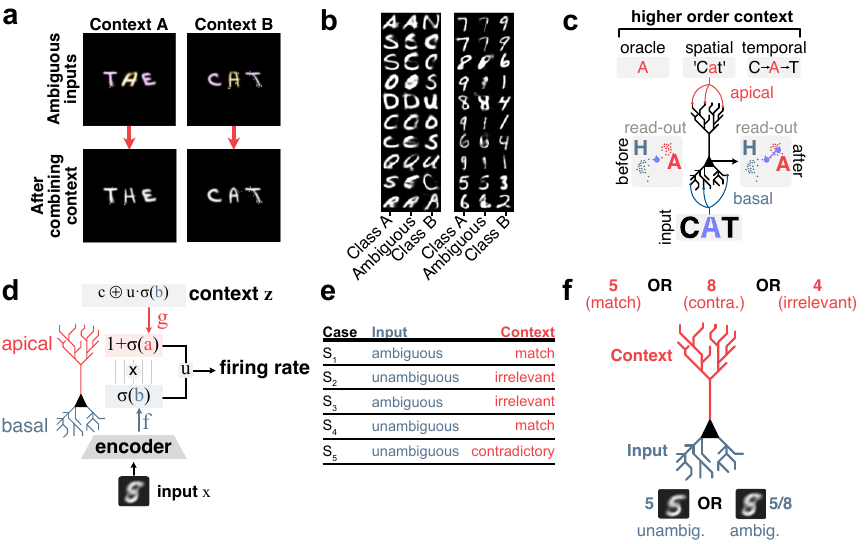}
\caption{\textbf{Functional model of context integration in apical dendrites}. 
\textbf{a}, in our experimental paradigm, ambiguous characters (in yellow, top panels) are resolved by combining context signals (bottom panels).
\textbf{b}, examples of unambiguous pairs and resulting ambiguous characters generated using EMNIST (left) and MNIST (right).
\textbf{c}, model of pyramidal neurons with distinct dendritic compartments trained to solve a multi-scenario contextual integration task with ambiguity. Before top-down modulation, two predictions have equal probability. After top-down modulation, context contributes to favoring one unambiguous representation.
\textbf{d}, implementation of dendritic integration in artificial neurons with apical and basal compartments. $f$ represents the feedforward mapping onto the basal compartment parameterized by the pre-trained backbone, and $g$ the contextual top-down mapping onto the apical compartment, parameterized by its own neural network. The combination of the outputs $b$ and $a$ of $f$ and $g$, respectively, is applied neuron-wise via the multiplicative interaction $\sigma(b) * (1 + \sigma(a))$.
\textbf{e}, distinct combinations of input and context used to train the top-down network $g$.
\textbf{f}, possible combinations of input and context used during training and analysis.
}
\label{fig:fig1}
\end{figure}

To model how apical dendrites integrate context to refine perceptual representations, we first developed a multi-scenario contextual integration task where correct classification of ambiguous characters requires appropriate integration of contextual cues (Fig.~\ref{fig:fig1}\textbf{a}).
To create the task, we trained a generative model on handwritten digits and letters (MNIST and EMNIST, respectively) to construct a dataset with highly ambiguous characters. We used the generative model to synthesize images that are a mixture of two classes (see Methods). The ambiguity of the synthesized images was empirically quantified via softmax on classifier predictions, where we select images near 50\% for two classes
(fig.~\ref{fig:fig1}\textbf{b}).
Inspired by anatomical features of pyramidal neurons, we next developed a model that implements distinct dendritic compartments, a basal compartment for integrating sensory information and an apical compartment for integrating contextual information (Fig.~\ref{fig:fig1}\textbf{c}).
In the model, a population of $n$ pyramidal neurons has two sources of inputs, represented as vectors: a set of basal inputs, $\mathbf{b} = [b_1, \hdots, b_n]$, representing the sensory stream, and a set of apical inputs, $\mathbf{a} = [a_1, \hdots, a_n]$, representing the top-down stream.
Input images, $\mathbf{x}$, are encoded by a pre-trained network $f$, with parameters $\Theta_b$, where
$f$ is a convolutional neural network reflecting initial stages of sensory processing (see Methods for more details on how the encoder is pre-trained).
The output of $f(x)$ is the basal vector $\mathbf{b}$.

The firing rate of pyramidal neurons, $\mathbf{h} = [h_1, \hdots, h_n]$, is then determined by a thresholded neuron-wise (element-wise) combination of the basal and apical input vectors:

\begin{equation}
\mathbf{h} = \sigma(\mathbf{b}) \odot (\sigma(\mathbf{a})+1),
\label{eq:soma_output}
\end{equation}

where $\odot$ is an neuron-wise multiplication (Hadamard) and $\sigma$ is the rectified linear unit (ReLU) activation function implementing basal and apical non-linearities, respectively.
As can be seen from equation \ref{eq:soma_output}, the impact of the apical input vector, $\mathbf{a}$, on the firing rate, $\mathbf{h}$, is determined by $\sigma(\mathbf{a})+1$.
As such, the apical activity serves as a thresholded neuron-wise gain modulator of the basal activity, in-line with previous experimental reports of a multiplicative role for apical inputs \citep{waters_supralinear_2003}, an architectural bias of neocortical networks we refer to here as the "apical prior." 
Under these constraints, apical inputs affect neuronal activity only when they exceed their activation threshold (when $a > 0$), analogous to observations that apical inputs influence somatic spiking only if they evoke a dendritic calcium spike \citep{larkum_new_1999}. Thus, this neuron-wise gain modulation of basal representations by the apical compartment via the Hadamard product is an implementation of the "apical prior."  

To monitor the ability of the model to solve the perceptual task, we read-out model predictions from a linear transformation of the pyramidal neurons firing rates:

\begin{equation}
\label{eq:mu_out}
\mathbf{\mu} = \mathbf{U} \mathbf{h},
\end{equation}

where $\mu$ represents a higher-order neuronal population and $\mathbf{U}$ is a linear transformation (to the latent space of the encoder $f$).
To obtain the apical inputs $\mathbf{a}$, the context signal, $\mathbf{c}$, is concatenated with $\mu$, and mapped onto the apical compartment using a multi-layer perceptron (MLP), $g(\cdot)$, as follows:

\begin{equation}
\mathbf{a} = g(\mathbf{c} \oplus \mathbf{U\sigma{(b)}}),
\label{eq:apical_compartment}
\end{equation}
where $\oplus$ is the concatenation operation, $\sigma$ is the ReLU activation, and $g$ is a MLP with 1 hidden layer, and its parameters are denoted by $\Theta_a$.

As a population, these pyramidal neurons integrate sensory inputs (representations of characters) onto their basal dendritic compartment on one hand, and contextual information onto their apical dendritic compartment on the other hand.

For ambiguous characters, which are a mixture of two categories, a readout of the representations of the model will yield two plausible interpretations, and contextual information is required to resolve the ambiguity. We consider the category which is compatible with the contextual input to be the target (match) category. Conversely, a category which is a plausible interpretation of the input but not compatible with the contextual information is referred to as the contradictory category. We refer to all other categories as irrelevant categories. In total, we consider 5 possible scenarios. The labels and corresponding descriptions we use for each scenario are provided explicitly below:

\begin{itemize}
    \item \textbf{ambig.Match}: ambiguous input, helpful (matching) context
    \item \textbf{unambig.Irrel}: unambiguous input, irrelevant context
    \item \textbf{ambig.Irrel}: ambiguous input, irrelevant context
    \item \textbf{unambig.Match}: unambiguous input, helpful (matching) context
    \item \textbf{unambig.Contra}: unambiguous input, contradictory context
\end{itemize}

To test whether the contextual integration model incorporating the "apical prior" can learn apical modulations that are helpful for resolving ambiguous stimuli, we trained the synaptic weights $\Theta_a$ with the objective of modulating basal representations to match the target $\mu$ (representation after linear transform). We optimize this objective simultaneously for the set of all input-context scenarios listed above (Fig.~\ref{fig:fig1}\textbf{e} and Eq.\ref{eq:lossfn}) using a gradient descent optimization algorithm (Adam, see pseudocode \ref{alg1}). 

This setup ensures that the top-down model is trained on both unambiguous and ambiguous examples, and that the model must learn to provide the appropriate modulatory signal only when contextual inputs provide useful information for solving the task, and ignore them when they are irrelevant or contradictory to sensory information. Our proposed implementation reflects the reality that, in the mammalian brain, top-down modulation is available from multiple sources regardless of the ambiguity of the sensory inputs or the relevance of the contextual information. 

This suggests an approach where the problem of contextual integration is broken into two parts:
(1) how apical dendrites locally learn to use contextual representations $c$ to solve the multi-scenario task and (2) how useful contextual representations $c$ can be learned from available sources (e.g. temporal information).

For the first part, our initial experiments assume that context, $c$, is given in the form of a one-hot vector, a scenario we refer to as the 'oracle context' (see Methods).
Subsequent experiments for the second part address the problem of learning a representation of context, $c$, taking inspiration from the neocortex, where context is, in part, represented by higher-order regions with broader windows of 
temporal integration \citep{eichenbaum_integration_2017} (Fig.~\ref{fig:fig1}\textbf{c}; see Methods).

\subsection{Resolving ambiguity in the oracle context}
\begin{figure}[htp]
\includegraphics[width=\linewidth]{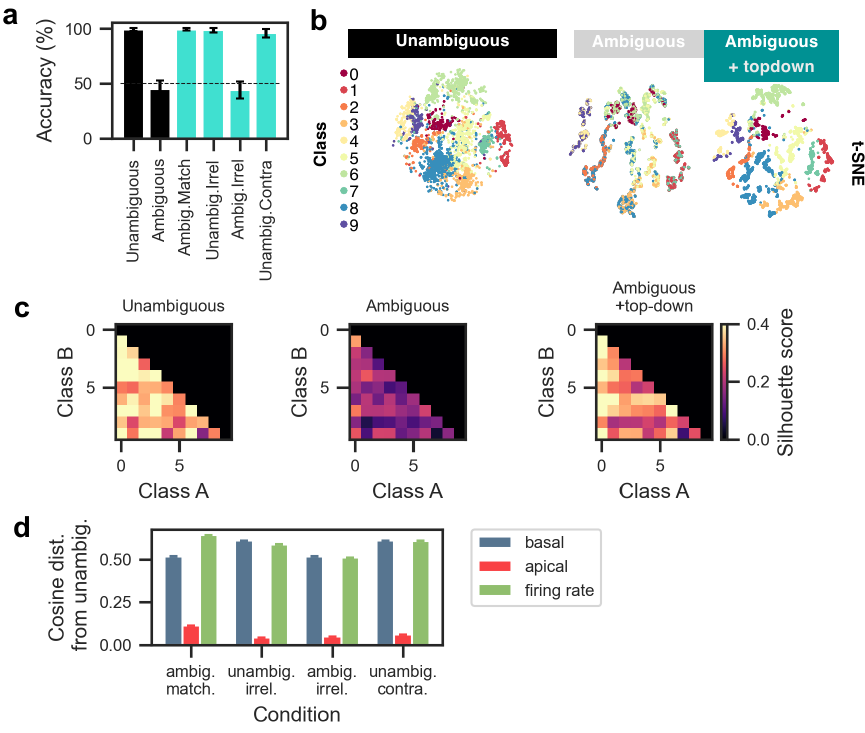}
\caption{\textbf{Our task is best solved when top-down signals are integrated correctly}.
\textbf{a}, readout test set accuracy without (black) and with (cyan)  top-down signals for all input-context combinations. Dashed line corresponds to chance level.
\textbf{b}, low dimensional projection (t-SNE) of the latent representation, $\mu$, for unambiguous inputs (top), ambiguous inputs before (bottom left), and after (bottom right) top-down modulation.
\textbf{c}, silhouette scores for latent representation, $\mu$, for each class pair and for unambiguous inputs (left), ambiguous inputs before (center), and after (right) top-down modulation.
\textbf{d}, cosine distance between unambiguous activation, $h$, and basal, apical compartments and firing rate in each scenario.
}
\label{fig:fig2}
\end{figure}

To first test if our model can solve our multi-scenario task under contextual input represented as a one-hot vector (oracle context), we trained $g$ on the ambiguous MNIST/EMNIST  task with a loss function that includes a loss term for each scenario
We assessed performance of the model in each scenario as the accuracy of the readout on the test set, and found significant impact of top-down modulation. 
(one-way ANOVA; $F_{4}$ = 10076.598; p $<$ 0.0001; n = 313 independent samples).
Specifically, test set accuracy on ambiguous digits was $46.3 \pm 0.3\%$ without top-down modulation, and became significantly higher with top-down modulation ($98.5 \pm 0.3\%$; Tukey's test; p $<$ 0.0001) if context was matching, but did not significantly alter performance if context was irrelevant ($43.920 \pm 8.160$; Tukey's test; p = 0.4651; n = 313 independent replicates; Fig.~\ref{fig:fig2}\textbf{a};  Table \ref{tab:results}). 
Importantly, the model did not learn to rely solely on top-down context, since performance remained high when the model was given contradictory context (Fig.~\ref{fig:fig2}\textbf{a}; \ref{tab:results}).
To visualize the representations of the model on a population level, we projected un-modulated and modulated outputs $\mathbf{\mu}$ onto a two-dimensional manifold using t-SNE (\cite{maaten_visualizing_2008}; Fig.~\ref{fig:fig2}\textbf{b}).
Ambiguous images whose representations exhibited high overlap between class-pairs were effectively disentangled after integrating matching contextual inputs (paired t-test, t$_{99}$ = 6.0293, p $<$ 0.0001 for silhouette scores between ambiguous images before and after integrating top-down inputs; Fig.~\ref{fig:fig2}\textbf{c}).
These results show that our model learned to use contextual inputs provided to apical dendrites to modulate basal activity and resolve ambiguity.

\begin{table}[]
\centering
\begin{adjustbox}{max width=\linewidth}
\begin{tabular}{@{}lllllllll@{}}
\toprule
Task & Context signal & ambig (Baseline) & ambig.Match & unambig.Match & unambig.Irrel & ambig.Irrel & unambig.Contra \\ 
\midrule
aMNIST & oracle (one-hot) & $46.3 \pm 0.3$ & $98.5 \pm 0.3$ & $98.5 \pm 0.1$ & $97.9 \pm 0.1$ & $46.0 \pm 0.8$ & $96.0 \pm 0.4$ \\
aEMNIST & oracle (one-hot) & $46.5 \pm 0.5$ & $94.2 \pm 1.4$ & $98.9 \pm 0.2$ & $96.2 \pm 0.4$ & $47.1 \pm 0.8$ & $94.9 \pm 0.5$ \\
\bottomrule
\end{tabular}
\end{adjustbox}
\caption{Ambiguous MNIST and EMNIST (aMNIST, aEMNIST) test set classification accuracy across input-context scenarios with the Hadamard integration rule. Results are expressed for each scenario as mean \% $\pm$ standard deviation with 3 random seeds.}
\label{tab:complete_results}
\end{table}

Next, we evaluated the importance of having top-down apical compartments to handle contextual inputs compared with only having basal compartments (akin to point-wise spiking neurons, (Table. \ref{tab:mnist_ablations}). We found that, while there was not a large difference in the other 4 scenarios, there was an average increase in performance of
1.9\% (MNIST)
in the unambig.Contra scenario for the models which had apical compartments compared to those which only had basal compartments.

\begin{figure}[htp]
\includegraphics[width=\linewidth]{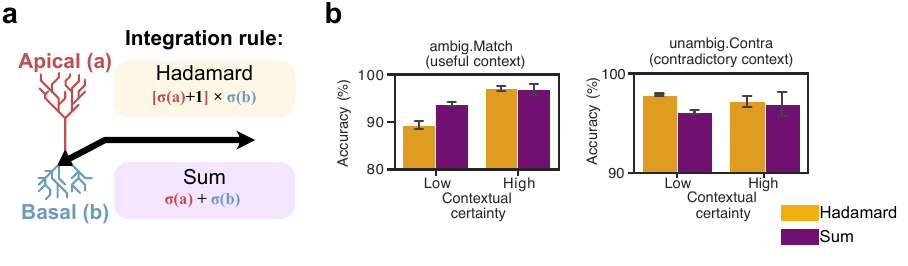}
\caption{\textbf{Role of integration rules on ambiguity resolution}. \textbf{a}, comparison between Hadamard and additive integration rules. \textbf{b}, test set accuracy relative to contextual certainty using either integration rule in conditions where inputs are ambiguous and context is relevant (ambig.Match) or where inputs are unambiguous and context is contradictory to sensory input (unambig.Contra).} 
\label{fig:results_integration}
\end{figure}

Next, focusing our analysis on MNIST, we evaluated the importance of the integration rule which combines basal and apical activities (Fig. \ref{fig:results_integration}\textbf{a}, Appendix 1 Table. \ref{tab:mnist_ablations}).

We found a significant interaction between the integration rule and contextual certainty
(2-ANOVA, F$_{1}$ = 5.4741, p = 0.0346). 
When inputs are ambiguous and the context is highly certain (context vector with values $>$ 0.99 for the target class), we find that the additive and multiplicative (Hadamard) integration rules perform similarly.  
(pairwise t-test, t$_{4}$ = 0.1708, p = 0.8678). 
For lower levels of certainty (context vector with values $<=$ 0.6 for the target class), we note the following observations: 1) when context is matching, Hadamard integration is outperformed by Sum integration
(pairwise t-test, t$_{4}$ = -4.7677, p = 0.0089, and 
2) when context is contradictory, Hadamard integration outperforms Sum integration (pairwise t-test, t$_{4}$ = 8.9341, p = 0.0009;
Fig. \ref{fig:results_integration}\textbf{b}).

\subsection{Single neuron mechanisms of top-down modulation}
We next sought to understand the computational role of top-down modulation of pyramidal apical dendrites in solving this contextual integration task.
To this end, we used a method from the explainable AI literature to rank and identify the neurons most relevant for solving the task.
We then compared their activations across the various context-input scenarios.
First, to define the most relevant neurons, we employed layer-wise relevance propagation (LRP; \citep{bach_pixel-wise_2015}; Appendix Fig.~\ref{fig:results_LRP_analysis}\textbf{a}), a method in the explainable AI literature to reveal, at each layer, the input's contribution to the model output. Each neuron receives a score for its contribution to the model output for each image (Appendix Fig.~\ref{fig:results_LRP_analysis}\textbf{b}).
We then defined the subset of neurons most relevant for predicting a given class, e.g. class A or class B, as the minimum set of neurons required to account for 95\% of total neuronal relevance when processing images from that class (Appendix Fig. ~\ref{fig:results_LRP_analysis}\textbf{c,d}, see Methods). Similarly, we define subsets of neurons most relevant for predicting that an image is ambiguous between two classes A and B (ambiguous AB).
Sets of relevant neurons defined in this way exhibit low overlap between classes (Appendix Fig.~\ref{fig:results_LRP_analysis}\textbf{b-d}, see Methods).
We observed similar representational separability between classes for other scenarios (Appendix Fig.~\ref{fig:results_LRP_analysis}\textbf{e}).
Additionally, these sets of relevant neurons are typically sparse and represent $\sim$15\% of the total neuronal population (Appendix Fig.~\ref{fig:results_LRP_analysis}\textbf{b,c}).

\begin{figure}[htp]
\includegraphics[width=\linewidth]{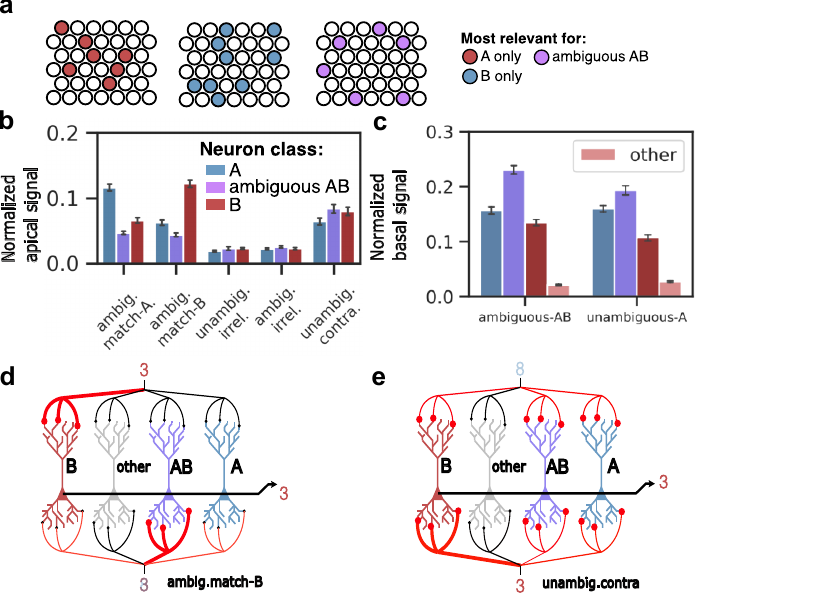}
\caption{\textbf{Top-down has information about ambiguity, and in cases where inputs are ambiguous and context is aligned, resolves ambiguity by amplifying context-relevant neurons}. 
\textbf{a}, using layer-wise relevance propagation, we identify neurons most relevant for either of two classes (red and blue), or for ambiguous inputs between the two classes (magenta).
\textbf{b}, amplitude of the apical compartment (normalized over all neurons) for each input/context condition. Three distinct classes of neurons are compared: neurons most relevant for unambiguous class A (blue), B (red), or neurons most relevant for the ambiguous AB images (magenta).
\textbf{c}, Normalized basal signal. left: in case of ambiguous stimuli, both plausible class-subpopulations are active along with the ambiguity detector subpopulation. right: in case of unambiguous stimuli, we observe a similar distribution, with the context-relevant neurons (class A) having higher average firing rate.
\textbf{d}, a cartoon summary of the observed mechanism for the ambig.match-B case (ambiguous input, contextual input favouring class B), where class B neurons code for '3' (dark red) and class A neurons code for '8' (blue). Context-relevant neurons (dark red) receive the highest amplitude of top-down signals when processing ambiguous input. At basal dendrites, these sub-populations have low but non-zero firing rates, which allows them to be selectively amplified with top-down signals via the Hadamard integration rule, and thus dictate the output representation.
\textbf{e}, for unambiguous input and contradictory context (unambig.contra), top-down input becomes non-specific and neurons are weakly amplified by the Hadamard rule. The output is then dictated by the neurons receiving the strongest basal input, in this case class B neurons (dark red). Output representations for unambiguous inputs are thus preserved when contextual inputs are added.} 
\label{fig:results_neuron_mechanisms}
\end{figure}

It is noteworthy that our model is not explicitly provided with the ambiguous nature of the input images, and therefore learns to extract this information to solve the task.
In addition, the model learns to ignore top-down signals when the image is unambiguous and context is contradictory or irrelevant.
Given that the apical compartment must combine contextual information and the contextually naive $\mu$ representation differentially based on input ambiguity, we hypothesized that solving the task effectively required a high gain modulation specific to the subset of context-relevant neurons.
To assess this possibility, we computed the average amplitude of apical signals arriving at LRP subsets pertaining to predicting each of two input classes (A,B) and ambiguous AB for each scenario of matching, irrelevant or contradictory context (Fig.~\ref{fig:results_neuron_mechanisms}\textbf{a-f}).
When comparing all input/context scenarios, we found that apical signals were highest when input images were ambiguous and context informative 
(1-ANOVA, F$_{416}$=237.024, p $<$ 0.0001, for the main effect of input/context scenario; 
Fig.~\ref{fig:results_neuron_mechanisms}\textbf{b}).
In the ambig.Match scenario, neurons most relevant for the target class were associated with the highest amplitude in the apical compartment 
(1-ANOVA, F$_{417}$=17.9445, p $<$ 0.0001, for the main effect of neuron class; 
Fig.~\ref{fig:results_neuron_mechanisms}\textbf{b}).
Apical input onto  neurons outside of these defined subpopulations ("other") remained low ($\sim$ 0.07-0.08) across all scenarios (data not shown), indicating there was no relationship between other neurons' apical signal and the scenario.

We also computed the basal amplitude for the same subsets of neurons defined above and found that there was a higher average amplitude on the matching relative to the contradictory subpopulation for unambiguous stimuli. (Fig.~\ref{fig:results_neuron_mechanisms}\textbf{c}). Likewise, there was a higher basal amplitude on the ambiguous subpopulation relative to other subpopulations for ambiguous inputs.
Basal input onto the "other" subpopulation was found to be significantly lower for ambig.match and unambig.contra scenarios, but similar in magnitude to the unambig/ambig.irrel scenarios ($<$0.02, Fig.~\ref{fig:results_neuron_mechanisms}\textbf{c}).  

These observations are indicative of a mechanism whereby ambiguous bottom-up representations are biased towards top-down representations by specific apical inputs Fig.~\ref{fig:results_neuron_mechanisms}\textbf{d}, whereas unambiguous bottom-up representations are preserved due to non-specific top-down modulation Fig.~\ref{fig:results_neuron_mechanisms}\textbf{e}.


To evaluate if specific top-down modulation is necessary for solving the ambig.match task, 
we applied a mask on the apical inputs to the corresponding subpopulations post-training, and assessed the model's ability to solve the task (Appendix 2-Fig.~\ref{fig:mask_analysis}).
Specifically, we compare test set accuracy when inputs are ambiguous and context is relevant, under the condition where context-relevant apical inputs are masked out (activations set to 0), compared to the condition where random apical inputs are masked.
We also evaluate the effect of mask size and found that masking the apical inputs to context-relevant neurons specifically led to a steeper degradation of accuracy as we increase the number of masked inputs. 
This highlights the key role of these neurons' apical compartments in integrating relevant contextual signals.

\subsection{Deriving context from temporal information}
\begin{figure}[htp]
\includegraphics[width=\linewidth]{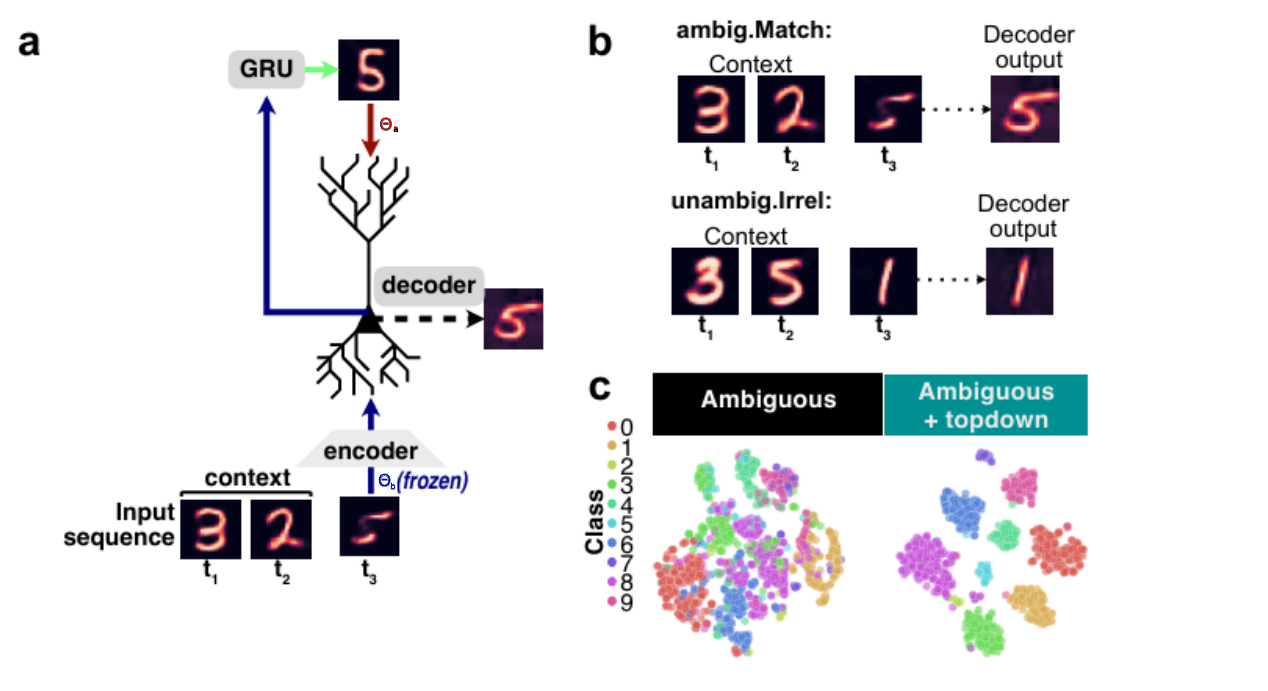}
\caption{
\textbf{Deriving context from temporal information}. \textbf{a}, rationale for training a model that leverages context derived from temporal sequences through a recurrent network (GRU). 
\textbf{b}, example sequence used to generate context signals (as the arithmetic sum of two digits, left and center columns) to resolve ambiguity of an input image (right column). 
\textbf{c}, t-SNE projection of latent representation associated with ambiguous inputs before (left) and after (right) combining top-down signals.}
\label{fig:results_temporal}
\end{figure}

Contextual priors can also be extracted from the temporal domain, specifically by leveraging past information to decode incoming ambiguous sensory inputs.
To extend our model to cases where contextual signals are derived from the temporal domain, we trained a Gated Recurrent Unit (GRU) network to predict the arithmetic sum of two unambiguous digits as encoded by our pre-trained feedforward weights $\mathbf{\Theta_b}$.
The final output state of the GRU was then given as a contextual signal for our contextual integration model.  (fig.~\ref{fig:results_temporal}\textbf{a}; see Methods).
We find that in cases where the input is ambiguous and the temporal sequence sums to a plausible interpretation of the ambiguous image, the model updates perceptual representations using contextual signals to successfully resolve ambiguities (fig.~\ref{fig:results_temporal}\textbf{b}, top).
Importantly, context is effectively ignored in cases where the input is unambiguous and the contextual signals are irrelevant (fig.~\ref{fig:results_temporal}\textbf{b}, bottom).
Using low-dimensional projections of latent representations, we find that top-down context derived from the temporal domain effectively disentangled overlapping representations when inputs are ambiguous, similar to the oracle case. (fig.~\ref{fig:results_temporal}\textbf{c})

\begin{table}[]
\centering
  \begin{adjustbox}{max width=\linewidth}
\begin{tabular}{@{}lllllll@{}}
\toprule
Task & Context signal & ambig.Match & unambig.Match & unambig.Irrel & ambig.Irrel & unambig.Contra  \\ \midrule
 seq-aMNIST  & temporal & 96.3 $\pm$ 0.1 & 99.9 $\pm$ 0.0  & 99.1 $\pm$  0.1 & 48.7 $\pm$ 0.5 & 99.0 $\pm$  0.1 \\
\end{tabular}
\end{adjustbox}
\caption{Sequential aMNIST (seq-aMNIST) test set classification accuracy across input-context scenarios with the Hadamard integration rule. Results are expressed for each scenario as mean \% $\pm$ standard deviation. In the temporal case, context signals are provided to the contextual integration network as GRU hidden state representations, conditioned on previous inputs in the sequence. Results are expressed for each scenario as mean \% $\pm$ standard deviation with 3 random seeds.}
\label{tab:results}
\end{table}

\section{Discussion}
 
The ability to appropriately use and switch contexts to make sense of a vast stream of sensory information is an important feature of robust cognitive systems.

It is thought that in the neocortex, contextual representations in higher order regions modulate lower sensory regions through top-down interactions arriving on the apical dendrites of pyramidal neurons \citep{larkum_cellular_2013, phillips_cognitive_2017}, an architectual bias we refer to here as the "apical prior."
While it is known that apical dendrites modulate neuronal firing, their specific computational role in cognitive processes such as contextual integration remains poorly understood. 



Here, we tested whether the "apical prior" represents an architectural bias that is functionally useful for contextual integration in neural networks. To this end, we implemented (1) a neural network architecture with distinct apical compartments and a neuron-wise integration rule based on the observation that apical dendrites are gain modulators of somatic activity \citep{waters_supralinear_2003, larkum_dendritic_2007}, and developed (2) an ambiguous image classification task that requires contextual integration to be solved.
We trained (1) the contextual modulator (top-down synapses onto apical dendrites) by gradient descent to simultaneously learn all input-context conditions defined in our task.
In this setting, we found that this bio-inspired dendritic architecture outperformed a single-compartment model which processes both sensory and contextual information with a single non-linear activation.


By applying LRP, we analyzed the amplitude of apical inputs onto class-specific neurons for ambiguous stimuli. We found the highest gain modulation in context-relevant neurons (matching the target class), consistent with the principles of biased competition \citep{spratling_predictive_2008}.

Moreover, we found that the competition introduced by the contextual integration adapted to the level of ambiguity in the task: The top-down network learned to a) refine representations of ambiguous stimuli by applying strong but sparse gain modulation to basal signals when contextual information is both available and relevant, and b) provide weak and non-specific modulation in the case of unambiguous stimuli and contradictory or irrelevant contexts, thus preserving the bottom-up representations. 
%

\subsection{Comparison with existing computational frameworks}


From a computational perspective, apical dendrites have been previously shown to support some types of contextual integration in biophysically realistic neurons and ANNs \citep{naumann_invariant_2022, wybo_dendritic_2022, george_detailed_2020, baronig_context_2024}. The tasks used in these previous works are multi-task, single scenario problem settings, meaning that context (e.g. one-hot encoded task id) is always assumed to be relevant. 
Alternative approaches to flexible contextual integration have involved explicitly learning to prevent associations between irrelevant context and sensory inputs \citep{baronig_context_2024}, which represents a successful example of how to handle spurious contexts in a simple task.
As an extension of this approach, our model must infer how to handle the interaction between sensory input and context across distinct tasks and scenarios, where context is relevant, irrelevant, or contradictory, but without explicit knowledge of the task.

Crucially, we only found a significant advantage of the apical prior over single compartment integration in the scenario where inputs are unambiguous and context is contradictory (and should be ignored). This highlights a specific advantage for apical dendrites in the role of contextual integration, where accounting for varying contextual relevance is crucial to broadly robust performance. Furthermore, while we show that the task can be solved using either a gain modulation (Hadamard) or additive apical integration rule (Appendix 1 Table. \ref{tab:mnist_ablations}), we found that, under low contextual certainty, the Hadamard integration rule (apical prior) enabled the model to ignore contradictory contextual signals more effectively compared to additive apical integration. 

Another key difference of our study with most previous work is that as basal and apical signals are combined post-activation, our rule can only amplify already active neurons, unlike \cite{wybo_dendritic_2022} and \citep{naumann_invariant_2022}, but inline with experimental evidence \citep{larkum_new_1999}. 
Thus, our approach maintains the sparsity of the network representations, which can preserve selectivity, and would also be more metabolically efficient.

One notable prior work which applies context-dependent modulation to feedforward networks is context-dependent gating (XdG) \citep{masse_alleviating_2018}, which implements contextual inputs by concatenation of context signals and gating task-specific neurons. The former could be comparable with our single compartment (basal-only) ablation as it employs a single zone of integration.
The main difference with our basal-only ablation is that XdG a) is not informed by intermediate sensory representations, and b) as most prior works, assumes task labels are given and that context always agrees with sensory input. Therefore, it is unknown if XdG, implemented as-is, can deal with cases where context should be overridden by sensory input, as in the unambig.Contra case.

In our experiments, the advantage of apical dendrites was revealed in our multi-scenario contextual integration task. In order to compare our results with previous work, we examined the effect of our integration mechanism in a standard multi-task setting (with task-ids provided to the network, \ref{tab:mnist_ablations}). Under these conditions we found that all variants perform similarly well. This can be explained by the setting being simpler, whereas the distinct apical compartment shows its advantage under more varied and challenging types of contextual integration that require inference of ambiguity (no task ids provided). 

While not explicitly modular, our framework could be incorporated into the Global Workspace Theory (see \citep{goyal_coordination_2022} for a comparison of recent implementations), to provide a candidate mechanism for resolving inconsistency between modules under ambiguity.
It is important to note that our architecture embodying the apical prior presupposes there is asymmetry for how contextual (top-down) and perceptual (bottom-up) representations are combined under ambiguity, which is included in some (e.g. \citet{mittal_learning_2020}) but not other \citet{goyal_coordination_2022} implementations of Global Workspace Theory.
In general, modular representations can be learned in a distributed and scalable manner, with interactions between local populations (or modules). As highlighted in \citet{goyal_coordination_2022}, this frees up the capacity of a learning system for more tasks, as it can more efficiently identify task-relevant information and reuse what it has learned. The "apical prior" here implies that these interactions have an asymmetry: they can be contextual signals (in our case via top-down projections to apical compartments), or feed-forward/bottom-up signals. Here we find that this asymmetry provides flexibility to solve perceptual tasks under diverse ambiguity and contextual scenarios. 




\subsection{What is necessary for robust contextual integration?}


Thus far, we have shown that in our contextual integration task, the apical prior provides a specific advantage compared to single-compartment integration when sensory or contextual signals are unreliable. 
Here, we extend these findings to make broader predictions for what specific components/computations might be necessary for robust contextual integration, and how such a mechanism could plausibly be implemented in the cortex.
Our results suggest that it is sub-optimal to sum all activity together in a single (basal) compartment, if sensory information and context can be contradictory, as would be the case for general cognitive systems.
This highlights a key difference between single and two-zone integration of information. As such, an interesting experiment would be to compare the single vs two-zone integration of bottom-up representations and top-down contextual representations in in-vivo experiments or simulations of neurons which more faithfully respect the biophysical constraints of pyramidal neurons in the neocortex, to observe whether our results hold in a more biologically plausible setting, or if it is just a feature of ANNs and/or how they were trained here.


Furthermore, based on our results, we can infer that the apical compartment encodes some information about the ambiguity of the sensory input, and the relevance of context for a given sensory input. An interesting experiment could involve silencing or deactivating apical dendrites in more realistic neurons  (biophysical simulations or in-vivo experiments), to assess the effect on accurate perception for our proposed contextual integration tasks.  


Our analysis showed that mutual information between apical and basal activation 
is positive for the ambig.Match scenario. On the other hand, mutual information is close to 0 for the unambig.Irrel and unambig.Contra scenario (Appendix, Table 3).
This is not surprising given we explicitly trained under the assumption that local representations, given matching unambiguous contextual information, should collapse to the target representation.
Moreover, when sensory ambiguity is low, local representations should be invariant to contextual ambiguity, and mismatching contextual information. 
This is an important feature of the model supporting robust contextual integration 
given that most available contextual signals in rich, multi-modal real-world sensory experiences are distractors for any one particular task, and should be discarded.

Following this, we predict that a model cannot learn the relevance of context without encountering and recognizing cases with irrelevant context during training (negative samples). 
Extending this idea to unsupervised learning, it is interesting to think about how the model could identify contextual relevance. In an unsupervised setting this could arise from an information-maximization learning rule like InfoNCE, which uses a contrastive loss with positive and negative pairs defined via augmentation of the input (positive, relevant) against other inputs (negative, irrelevant). Unfortunately, the creation of positive and negative pairs relies heavily on modality-specific knowledge to construct useful data augmentations (e.g. a rotation applied on an image). A more general approach will be necessary to create such contrastive pairs for many modalities in parallel.


%

An important feature of our model is that it could learn a 
mapping of top-down signals onto the apical compartment separately from the contextual representations. This is compatible with regions high in the cortical hierarchy, such as the hippocampus (via the entorhinal cortex), with independent mechanisms for learning a) how they represent contexts, and b) how those contextual representations modulate the sensory representations of lower regions \citep{maren_contextual_2013}.
Because we assume the stability of the sensory and contextual representation prior to training contextual integration, we cannot make any conclusions from the present study about how these two representations could be simultaneously co-learned. 
However, we speculate that in the neocortex, there should be some degree of stability in the sensory and contextual representations before learning to perform contextual integration. 

This could be compatible with the distributed learning of modules (cortical regions) for sensory and contextual representations initially via self-supervised learning, prior to learning the contextual interactions between the modules (carried by cortico-cortical white matter projections).   

\subsection{Limitations}
While our model unifies the function of contextual integration with key biophysical properties of dendrites, some of our modeling assumptions and simplifications are worth discussing with respect to their empirical support.
Firstly, unlike other computational frameworks such as \citet{george_detailed_2020} which seek to incorporate the rich diversity and laminar architecture of the neocortex, our model is designed to implement some prominent features of cortical pyramidal neurons without focusing on a specific cortical layer, such as layer 2/3 or layer 5 pyramidal neurons.
Specifically, we focus on the functional properties of two-zone dendritic integration. In the cortical circuit, these mechanisms appear to be conserved across L2/3, L5, L6 pyramidal neurons \citep{larkum_new_1999, waters_supralinear_2003, larkum_dendritic_2007, ledergerber_properties_2010}.
Here, we assume that contextual signals are provided by neurons in a higher-order region, which would likely correspond to outputs of layer 5 pyramidal neurons in that region, and integrated in the apical dendrites of putative layer 2/3, 4 or 5 pyramidal neurons \citep{bastos_canonical_2012, schuman_neocortical_2021}.
It is worth noting that top-down projections may also innervate the basal compartment of layer 2/3 neurons, and the lower regions of L5 pyramidal neuron basal dendrites \citep{harris_hierarchical_2019}. 
Despite these missing anatomical features,  the framework developed here could serve to study the functional role of laminar architecture in the neocortex: Our study relies largely on the functional mechanism of two-zone integration in pyramidal neurons, which appears to be consistent across pyramidal neuron types, despite potential differences in the sources of the inputs they receive. 


Furthermore, our model only implements top-down modulation at a single level in the hierarchy. An obvious next step would be to develop a hierarchical generalization of the approach. While it is unknown how much extra improvement this would provide, evidence of top-down neuron masking (a less expressive form of modulation) at every layer has been shown to improve out-of-distribution robustness \citep{liu_gflowout_2023}.

One potential criticism is that we train here using gradient descent, for which concerns have been raised as to its biologically plausibility \citep{lillicrap_backpropagation_2020}.  
The aim of this study is not to find biological learning rules, but rather to understand contextual integration. We use gradient descent as a "best in class" general learning approach to find good solutions to our task, which allows us to analyze the learned representations and the respective contributions of apical and basal compartments. 

Our framework lacks the implementation of biological plasticity rules. We leave to future work the exploration of solutions obtained via more biologically plausible plasticity rules such as \citet{lee_difference_2015, meulemans_minimizing_2022}, compared to those we obtained via backprop. Indeed the analysis above suggests that information-maximization plasticity rules between basal and apical compartments could be a promising direction.

\subsection{Outlook}

Often in human brains, inputs from different modalities are processed simultaneously. As a future direction, we propose that our model could be extended to hierarchies of moduels of sensory modalities interacting with each other.
Deep learning models that implement multi-modal interactions are an active area of research \citep{radford_learning_2021, alayrac_flamingo_2022, mustafa_multimodal_2022}.
While they excel at multi-modal generation conditioned on one modality, they struggle with conflicting information or contradictions in mixed-modality data \citep{chen_llava-mole_2024}.

As state of the art deep learning models increase exponentially in size, many recent works focus on scalable, parameter-efficient fine-tuning for target tasks, a prominent example being low-rank adapters (LoRAs) \cite{hu_lora_2021, chen_llava-mole_2024}. While LoRAs use resources efficiently, they lack a contextual integration component, and as such, they could be complementary to the "apical prior" architecture studied here. A future direction could combine LoRAs with the apical prior for efficient contextual integration tasks at scale.
For example, given that in our task, only a small subset of associations represent matching context, we expect that the weight matrices mapping contextual representations onto apical dendrites could be effectively replaced by LoRAs. This has better scaling properties due to using a smaller fraction of total network capacity. 


Initially, stable and useful representations of context could be achieved with self-supervised learning. We speculate that the brain might employ both the apical prior and sparse weights to learn to handle the large pool of available sources of context efficiently. 
Toward Global Workspace Theory, we argue that a) there must be a mechanism for resolving ambiguity or discrepancy between modules before the step in which modules compete to write to the shared global workspace (as in \citet{goyal_coordination_2022}), and b) implementation of the "apical prior" satisfies this requirement. Thus, our present study provides an important step in understanding the process by which brain-wide networks of many modalities can coordinate and integrate information.


\section{Methods}

\subsection{Ambiguous dataset generation}

We developed a new visual contextual integration task composed of multiple scenarios that incorporate varying levels of ambiguity of sensory input and informativeness of contextual information.

To create the image dataset for our task, we used a digit-conditioned generative approach based on conditional variational autoencoders (CVAEs).
First, we train a CVAE on standard a image dataset, either MNIST for digits or EMNIST for characters \citep{doersch_tutorial_2021, sohn_learning_2015}. 
We then use the learned latent space along with a digit-conditioned input vector to generate images that are ambiguous between two digits by interpolating between the one-hot vectors of the two digits.

We then assess the ambiguous samples using a trained classifier on the original dataset (either MNIST or EMNIST), and discard images that are outside the $50\% (\pm 5\%)$ decision boundary between the two digits. We organize our dataset of images into triplets, where two images are unambiguous from digits $y_0, y_1$, and the third is generated by the CVAE to be ambiguous between $y_0, y_1$. This dataset format simplifies the implementation of our task setup.

\subsection{Contextual integration model architecture}
The contextual integration model  consists of three components: the pre-trained backbone that computes representations of input images, the readout network for classifying their respective representations, and the top-down network which modulates the intermediate representations from the pre-trained backbone based on contextual information.
The backbone is implemented as a convolutional variational autoencoder (VAE), and was pre-trained on unambiguous characters with the standard ELBO loss. VAEs are known to learn a smooth latent space, which we assume here to be an important ingredient to allow learning of top-down contextual modulations. 
Next, we freeze the VAE weights and train a readout (MLP) on the VAEs latent representations of images with a classification objective (cross entropy loss).
Finally, we freeze these components and train the top-down network in two contextual tasks, with oracle and temporal context respectively, as described in the next sections. 
The top-down network $g(\cdot)$ is implemented as a multi-layer perceptron (MLP) mapping the concatenated VAE latents and context vector to apical activations (eq.\,\ref{eq:apical_compartment}), which then modulate the VAEs intermediate representation. Various modulation functions (Hadamard, additive, concatenation) are explored in the main text.

\subsubsection{Oracle context}
In the oracle case, context signals are provided to the top-down network $g$ in the form of a one-hot vector:
\begin{equation}
\mathbbm{1}_y(c_i):=\begin{cases}
1, & \text{if $i = y$}.\\
0, & \text{otherwise}
\end{cases}
\label{one_hot_encoding}
\end{equation}
 $\forall i \in {1\dots C}$ where C is the number of classes, $|\mathbf{c}| = C$.


\subsubsection{Temporal context}
In the temporal case, the information is presented in a sequence, and preceding elements determine the contextual information available to the model at the current time $t$. Therefore, we leverage a Gated Recurrent Unit (GRU) network \citep{cho_learning_2014} to dynamically capture and represent the temporal dependencies in the sequence. We trained our GRU to predict the modulo 10 sum of MNIST digits in a sequence of two digits', reflecting the task in our sequential MNIST/EMNIST dataset. An important detail is that the input to the GRU at time $t$ is not the image $x_t$ , but rather the latent $\mathbf{Ub}_t$, encoded by the pre-trained backbone.
After processing a sequence of digits ($x_{t-2}, x_{t-1}$), the output state $o_t$ of the GRU was then used as a top-down context signal, in place of the one-hot context from the oracle case. We used a hidden state of size 128 for the GRU.
    
\subsection{Training}

\subsubsection{Loss function}
The top-down network $g(x, c|\Theta_a)$ is explicitly trained to minimize a loss function jointly across the following 5 conditions: ambiguous input with matching context (ambig.Match), unambiguous input with matching context (unambig.Match), unambiguous input with irrelevant context (unambig.Irrel), ambiguous input with irrelevant context (ambig.Irrel), and unambiguous input with contradictory context (unambig.Contra).
We use mean squared error as the loss criterion between the predicted (computed as in equation \ref{eq:mu_out}) and the target latent representations. Each mini-batch is sampled from dataset $D$, and training data for each condition are equally balanced in the mini-batch.
\begin{equation}
    \mathbf{L}(\Theta_a;D) = \mathbbm{E}_{((x_0,y_0),x_{ambig},(x_1,y_1))\sim D, c_{match},c_{irrel}\sim C} [ \sum_{s=1}^5 ||\mu_s^{*}-\hat{\mu_s}||_2^2 ]
    \label{eq:lossfn}
\end{equation}
$x_0,x_1$ are unambiguous images sampled from $D$.\\
The optimum of $\Theta_a$ is given by:
\begin{equation}
    \Theta^*_a = argmin_{\Theta_a} L(\Theta_a;D)
\end{equation}
Here, $\mu_s^{*}$ is the target representation, which is for each scenario and context:

\begin{equation}
\mu_s^{*} :=\begin{cases}
\mathbf{U}\mathbf{\sigma(b_{c_{match}})}, & \text{if $s\in \{1\}$} \\
\mathbf{U}\mathbf{\sigma(b_{unambig})}, & \text{if $s\in \{2,3\}$} \\
\mathbf{U}\mathbf{\sigma(b_{ambig})}, & \text{if $s \in \{4\}$} \\
\mathbf{U}\mathbf{\sigma(b_{c_{match}})}, & \text{if $s\in \{5\}$} 
\end{cases}
\label{eq:tgtbyscenario}
\end{equation}
Where $b_{c} = f(x_c|\Theta_b)$ and \\
$b_{unambig} = f([x_0,x_1]|\Theta_b)$. \\
Predicted latents are given by:
\begin{equation}
\hat{\mu}_s :=\begin{cases}
\mathbf{U}\mathbf{\sigma(b_{ambig})\odot(1+\sigma(a_{b_{ambig},c_{match}}))}, & \text{if $s\in \{1\}$} \\
\mathbf{U}\mathbf{\sigma(b_{unambig})\odot(1+\sigma(a_{b_{unambig},c_{match}}})), & \text{if $s\in \{2\}$} \\
\mathbf{U}\mathbf{\sigma(b_{unambig})\odot(1+\sigma(a_{b_{unambig},c_{irrel}}})), & \text{if $s\in \{3\}$} \\
\mathbf{U}\mathbf{\sigma(b_{ambig})\odot(1+\sigma(a_{b_{ambig},c_{irrel}}))}, & \text{if $s \in \{4\}$} \\
\mathbf{U}\mathbf{\sigma(b_{unambig})\odot(1+\sigma(a_{b_{unambig},c_{contra}}})), & \text{if $s\in \{5\}$} \\
\end{cases}
\label{eq:predictedlatents}
\end{equation}
Where $\mathbf{a_{b,c}} = {g(\mathbf{Ub,c}|\Theta_a)}$.

\begin{algorithm}
\caption{Training of the top-down model with oracle context}\label{alg1}
\begin{algorithmic}[1]
\Require Dataset $D$, learning rate $\lambda$, dimension of basal activation $d_b$, number of classes $C$, top-down input $\mathbb{R}^{d_b+C}$, output:  $\mathbb{R}^{d_b}$, bottom-up network (frozen) $f_{\Theta_b}$, latent projection (frozen) $\mathbf{U}$, top-down network $g(\dot|{\Theta_a})$, batch size B
\State $\mathbf{\Theta}_{a} \gets \texttt{init\_kaiming()}$
    \For {$\texttt{triplet mini-batch} (\mathbf{x_0},\mathbf{y_0}), \mathbf{x_{ambig}}, (\mathbf{x_1},\mathbf{y_1}) \texttt{ sampled from } D$}
        \State $y \gets cat(\mathbf{y_0}, \mathbf{y_1})$
        \State $x \gets cat(\mathbf{x_0,x_{ambig}, x_1})$
        \State $\mathbf{c_{match}}, \mathbf{c_{contra}} \gets random\_choice(y, size=(2,B)) \texttt{ \# Sample context from y without replacement}$
        \State $\mathbf{c_{irrel}} \gets random\_choice(range(C) \not\in y,size=B) \texttt{ \# Sample non-matching context uniformly}$
        \State $c \gets cat(c_{match},c_{irrel},c_{contra})$
        \State $b \gets f(x) \texttt{ \# first pass, will be used later}$
        \State $\mu = \mathbf{U}b$
        \State
        $\mu c \gets \mu \oplus c $
        \State 
        $a \gets g(\mu c ; \Theta_a) $
        \State
        $\mathbf{\hat{\mu}} \gets \mathbf{U}(\textbf{relu}(b) \odot (1 + \textbf{relu}(a))) \texttt{  \# predicted latent for each scenario, see eq.\ref{eq:predictedlatents}}$
        \State
        $Loss \gets \textbf{mse}(\mathbf{\hat{\mu}}, \mu^*) \texttt{ \# target $\mu^{*}$ specified by eq.\ref{eq:tgtbyscenario}}$
        \State $\texttt{\# autograd optimizer step}$
        \State $\Theta_a \gets \Theta_a + \lambda \nabla_{\Theta_a} Loss$
      \EndFor
\end{algorithmic}
\end{algorithm}

\subsubsection{Layer-wise Relevance Propagation Analysis}
We applied layer-wise relevance propagation (LRP) to identify specific neurons whose activity, as described by $\mathbf{h}$ in our model, contributed most to the category readouts. For this analysis, we trained a readout to classify the standard categories (ie. digits 0-9) as well as all possible ambiguous pairs (ie. 3/5, 5/8, 1/7, etc.), which for MNIST gives a total number of categories $C=55$. To identify the subset of neurons that are most relevant for a given class $i$, we sorted neurons by highest to lowest average relevance per class.
We then computed the normalized cumulative sum of relevance of neurons, and define $S_i$ as the minimum set of neurons required to obtained 95\% of the total LRP for each class $i$ where $i \in {0,1, \cdots, C}$  (fig.~\ref{fig:results_LRP_analysis}\textbf{d}).
We computed the separability between sets $S_{i}$, $S_{j}$ as $1-|S_{i} \cap S_{j}| / |S_{i} \cup S_{j}|$ (Fig.~\ref{fig:results_LRP_analysis}\textbf{e}). 

\subsection{Statistical analysis}
Parametric tests where used when data distribution was normal and variance homogeneous, otherwise non-parametric tests were used and reported when appropriate.
\textbf{Meaning of abbreviated statistical terms}. 1-ANOVA: one-way ANOVA; 2-ANOVA, two-way ANOVA; RM-ANOVA, repeated-measure ANOVA.
Error bars and bands represent standard error of the mean, unless stated otherwise.
*, p $<$ 0.05; **, p $<$ 0.01; ***, p $<$ 0.001; ****, p $<$ 0.0001; n.s., not significant.

\subsection{Code availability}
All experiments were carried out using PyTorch.
We use the MNIST and EMNIST datasets,
where number of output classes are $C = 10$ for MNIST, and $C = 26$ for EMNIST.
The code for models and data analysis is publicly available under: 
https://github.com/ABL-Lab/expectation-clamp
The code used to generate the ambiguous datasets as well as links to pre-generated datasets are available under:
https://github.com/ABL-Lab/ambiguous-dataset
\section{Acknowledgements}
We thank Roberto Araya, Doina Precup, Irina Rish, and Yoshua Bengio for helpful discussions.
This study was supported by funding from the Institute for Data Valorization (IVADO), the Unifying Neuroscience and Artificial Intelligence - Québec (UNIQUE) research center, the CHU Sainte-Justine Research Center (CHUSJRC), Fonds de Recherche du Québec–Santé (FRQS), the Canada CIFAR AI Chairs Program, the Quebec Institute for Artificial Intelligence (Mila), and Google. B.A.R. was supported by NSERC (Discovery Grant: RGPIN-2020-05105; Discovery Accelerator Supplement: RGPAS-2020-00031), CIFAR (Canada AI Chair; Learning in Machine and Brains Fellowship) and the Canada First Research Excellence Fund (CFREF Competition 2, 2015-2016) awarded to the Healthy Brains, Healthy Lives initiative at McGill University, through the Helmholtz International BigBrain Analytics and Learning Laboratory (HIBALL).
Compute infrastructure was supported through a grant of computing time to E.B.M. from the Digital Research Alliance of Canada.
N.I. received additional support from a NSERC Undergraduate student research award, and IVADO and UNIQUE excellence fellowships.
M.T received additional support from the HBHL Graduate Fellowship.
B.T.G. received additional support from a UNIQUE excellence fellowship.

\bibliography{refs_zotero_050424}


\appendix
\begin{appendixbox}
\newpage
\renewcommand{\thefigure}{1}
\begin{figure}[H]
\includegraphics[width=\linewidth]{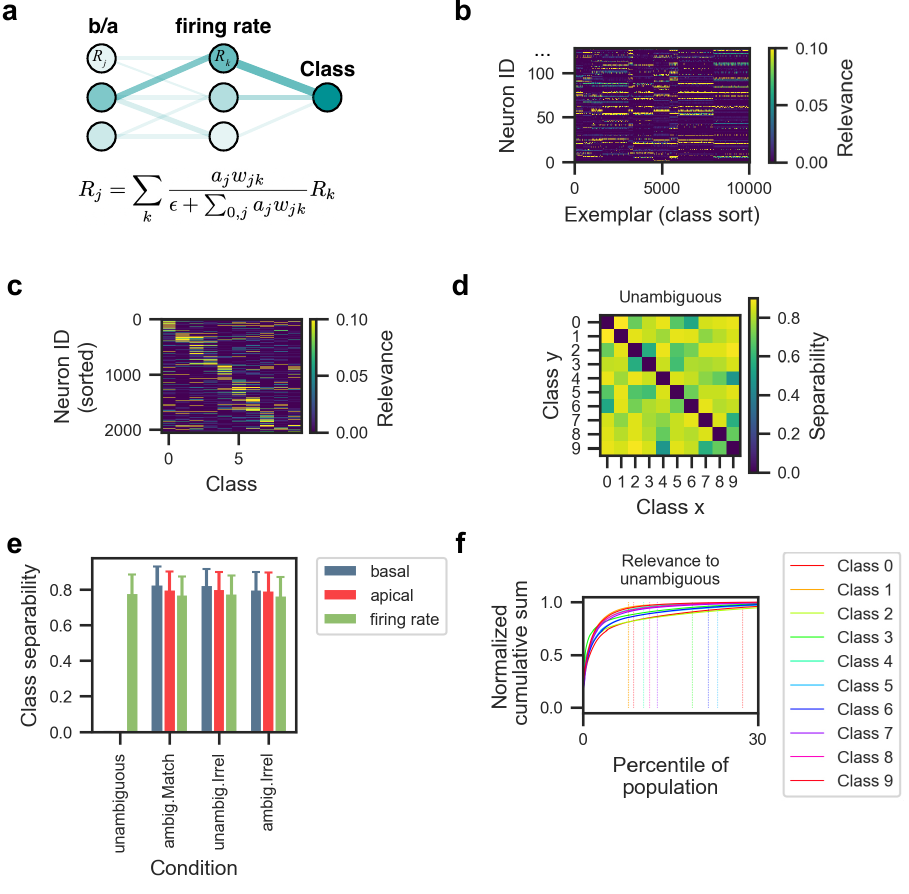}
\caption{\textbf{Rationale for identification of single neuron groups}. \textbf{a}, we computed relevance of neurons and their compartments (basal, apical, firing rate) using LRP.
\textbf{b}, Relevance of each neuron (soma) for input images sorted by class label.
\textbf{c}, average relevance for each neurons (ascending sort) for each class.
\textbf{d}, representational separability for every pair of classes.
\textbf{e}, class separability for each training scenario, and per dendritic compartment (blue, basal; red, apical; green, firing rate).
\textbf{f}, normalized cumulative sum of relevance for ranked neurons and for each class. Vertical dashed lines indicate the minimum number of neurons required to preserve 95\% of all relevance for each class.} 
\label{fig:results_LRP_analysis}
\end{figure}
\end{appendixbox}
    \begin{table}[]
\centering
\begin{adjustbox}{max width=\linewidth}
\begin{tabular}{@{}ccccccccc@{}}
\toprule
modulation & context\_only & backbone & scenario\_labels & ambig.Match & unambig.Match & unambig.Irrel & ambig.Irrel & unambig.Contra \\ \midrule
Hadamard   & no            & ae       & no              & 94.9 $\pm$ 0.8  & 98.4 $\pm$ 0.2  & 95.7 $\pm$ 0.1  & 46.9 $\pm$ 1.2  & 95.0 $\pm$ 0.1  \\
\midrule
Hadamard
   & no            & vae      & no              & 94.2 $\pm$ 1.4  & 98.9 $\pm$ 0.2  & 96.2 $\pm$ 0.4  & 47.1 $\pm$ 0.8  & 94.9 $\pm$ 0.5  \\
   & no            & vae      & yes             & 94.7 $\pm$ 0.6  & 99.2 $\pm$ 0.2  & 96.7 $\pm$ 0.2  & 47.2 $\pm$ 0.8  & 97.9 $\pm$ 0.2  \\
   & yes           & vae      & no              & 67.2 $\pm$ 0.8  & 97.3 $\pm$ 0.8  & 95.0 $\pm$ 0.5  & 47.6 $\pm$ 0.2  & 91.8 $\pm$ 1.7  \\
   & yes           & vae      & yes             & 76.4 $\pm$ 1.8  & 99.2 $\pm$ 0.1  & 96.5 $\pm$ 0.4  & 47.7 $\pm$ 0.4  & 97.9 $\pm$ 0.3  \\
\midrule
sum        & no            & vae      & no              & 99.1 $\pm$ 0.0  & 98.9 $\pm$ 0.3  & 96.2 $\pm$ 0.2  & 47.2 $\pm$ 0.6  & 94.3 $\pm$ 0.6  \\
        & no            & vae      & yes             & 99.7 $\pm$ 0.1  & 99.3 $\pm$ 0.2  & 96.4 $\pm$ 0.2  & 47.2 $\pm$ 1.3  & 97.8 $\pm$ 0.2  \\
        & yes           & vae      & no              & 79.3 $\pm$ 1.3  & 97.9 $\pm$ 0.4  & 95.3 $\pm$ 0.4  & 47.3 $\pm$ 0.6  & 92.3 $\pm$ 1.3  \\
        & yes           & vae      & yes             & 86.8 $\pm$ 0.5  & 99.1 $\pm$ 0.2  & 96.6 $\pm$ 0.1  & 47.6 $\pm$ 0.4  & 97.9 $\pm$ 0.2  \\
\midrule
concat     & no            & vae      & no              & 97.7 $\pm$ 0.8  & 99.0 $\pm$ 0.1  & 96.4 $\pm$ 0.2  & 47.1 $\pm$ 0.3  & 95.1 $\pm$ 0.6  \\
     & no            & vae      & yes             & 98.9 $\pm$ 0.7  & 99.3 $\pm$ 0.1  & 96.6 $\pm$ 0.1  & 47.6 $\pm$ 0.5  & 98.0 $\pm$ 0.0  \\
     & yes           & vae      & no              & 78.6 $\pm$ 1.6  & 98.1 $\pm$ 0.4  & 95.6 $\pm$ 0.1  & 47.1 $\pm$ 0.7  & 93.2 $\pm$ 0.1  \\
     & yes           & vae      & yes             & 84.9 $\pm$ 1.2  & 99.2 $\pm$ 0.1  & 96.5 $\pm$ 0.1  & 46.9 $\pm$ 0.3  & 97.9 $\pm$ 0.1  \\ \bottomrule
\end{tabular}
\end{adjustbox}

\caption{EMNIST test set accuracy across various metrics with different context representations, modulation strategies, and scenario labels. Results are expressed for each scenario as mean \% $\pm$ standard deviation with 3 random seeds.}
\label{tab:emnist_ablations}
\end{table}

\begin{table}[h!]
\centering
\begin{adjustbox}{max width=\linewidth}
\begin{tabular}{@{}ccccccccc@{}}
\toprule
modulation & context\_only & backbone & scenario\_labels & ambig.Match & unambig.Match & unambig.Irrel & ambig.Irrel & unambig.Contra \\ \midrule
Hadamard   & no            & ae       & no              & 98.4 $\pm$ 0.7  & 99.1 $\pm$ 0.1  & 98.3 $\pm$ 0.2  & 47.8 $\pm$ 0.2  & 97.3 $\pm$ 0.2  \\
\midrule
Hadamard   & no            & vae      & no              & 98.5 $\pm$ 0.3  & 98.5 $\pm$ 0.1  & 97.9 $\pm$ 0.1  & 46.0 $\pm$ 0.8  & 96.0 $\pm$ 0.4  \\
   & no            & vae      & yes             & 99.0 $\pm$ 0.4  & 99.7 $\pm$ 0.0  & 99.1 $\pm$ 0.1  & 46.3 $\pm$ 0.9  & 99.4 $\pm$ 0.2  \\
   & yes           & vae      & no              & 86.8 $\pm$ 1.0  & 94.0 $\pm$ 0.5  & 92.6 $\pm$ 0.7  & 46.8 $\pm$ 0.5  & 85.7 $\pm$ 0.9  \\
   & yes           & vae      & yes             & 91.7 $\pm$ 1.2  & 99.6 $\pm$ 0.1  & 99.1 $\pm$ 0.1  & 45.4 $\pm$ 0.7  & 99.4 $\pm$ 0.2  \\
   \midrule
sum        & no            & vae      & no              & 99.6 $\pm$ 0.1  & 98.1 $\pm$ 0.3  & 97.4 $\pm$ 0.3  & 46.2 $\pm$ 0.6  & 94.7 $\pm$ 0.6  \\
        & no            & vae      & yes             & 99.9 $\pm$ 0.0  & 99.5 $\pm$ 0.1  & 99.0 $\pm$ 0.0  & 46.7 $\pm$ 0.1  & 99.2 $\pm$ 0.1  \\
        & yes           & vae      & no              & 82.6 $\pm$ 4.0  & 96.2 $\pm$ 1.7  & 95.7 $\pm$ 1.6  & 47.1 $\pm$ 0.8  & 93.4 $\pm$ 2.3  \\
        & yes           & vae      & yes             & 89.6 $\pm$ 2.5  & 99.6 $\pm$ 0.2  & 99.1 $\pm$ 0.2  & 46.2 $\pm$ 0.2  & 99.4 $\pm$ 0.2  \\
\midrule
concat     & no            & vae      & no              & 98.8 $\pm$ 0.5  & 98.1 $\pm$ 0.2  & 97.4 $\pm$ 0.4  & 46.0 $\pm$ 0.4  & 94.1 $\pm$ 0.6  \\
     & no            & vae      & yes             & 99.2 $\pm$ 1.1  & 99.6 $\pm$ 0.1  & 99.1 $\pm$ 0.2  & 47.1 $\pm$ 0.4  & 99.4 $\pm$ 0.2  \\
     & yes           & vae      & no              & 81.1 $\pm$ 1.2  & 96.7 $\pm$ 1.0  & 95.9 $\pm$ 1.3  & 46.7 $\pm$ 0.3  & 93.8 $\pm$ 2.2  \\
     & yes           & vae      & yes             & 87.5 $\pm$ 0.8  & 99.7 $\pm$ 0.1  & 99.1 $\pm$ 0.1  & 46.7 $\pm$ 0.4  & 99.4 $\pm$ 0.1  \\ \bottomrule
\end{tabular}
\end{adjustbox}

\caption{MNIST test set accuracy across various metrics with different context representations, modulation strategies, and scenario labels. Results are expressed for each scenario as mean \% $\pm$ standard deviation with 3 random seeds.}
\label{tab:mnist_ablations}
\end{table}

\begin{table}[h!]
    \centering
    \resizebox{0.5\textwidth}{!}{
        \begin{tabular}{l l r r}
            \hline
            & & \multicolumn{2}{c}{\textbf{Mutual Information*}} \\
            \cline{3-4}
            \textbf{Scenario} & \textbf{Modulation} & \textbf{Mean} & \textbf{Std} \\
            \hline
            \textbf{ambig.Match} & \textbf{Hadamard} & 0.114 & 0.017 \\
            & \textbf{sum} & 0.107 & 0.020 \\
            \textbf{unambig.Contra} & \textbf{Hadamard} & -0.004 & 0.004 \\
            & \textbf{sum} & -0.004 & 0.005 \\
            \hline
        \end{tabular}
    }
    \caption{Mutual Information metrics for ambig.Match and unambig.Contra with Hadamard and sum operations. *We report MI as an approximate measure. This is computed by the expected information gain (which is equivalent to MI) on the readout predictions with and without top-down modulation given context. Since our readout prediction is a non-linear mapping from the apical and basal compartments, this is an approximate measure, as MI is only invariant to linear transformations.}
\end{table}
\begin{appendixbox}

    \renewcommand{\thefigure}{1}
\begin{figure}[H]
        \centering
        \includegraphics[width=0.5\linewidth]{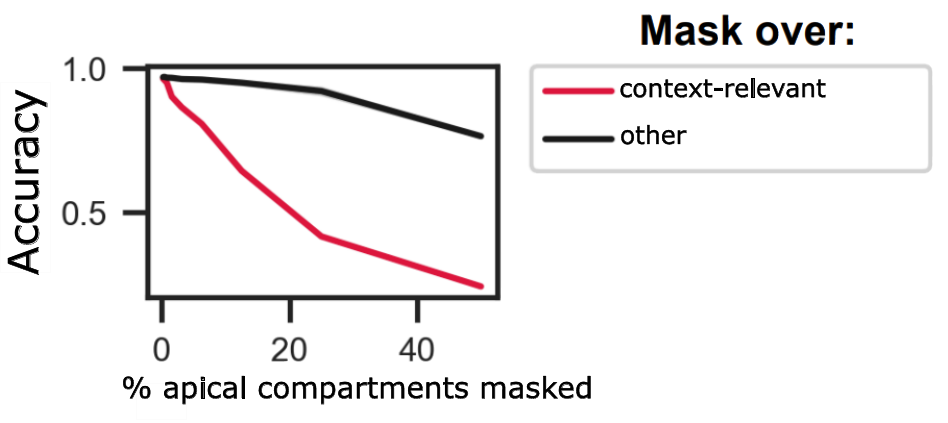}
        \caption{\textbf{Masking analysis.} 
        We compare test set accuracy when inputs are ambiguous and context is relevant (ambig.Match) with context-relevant apical inputs are masked out (activations set to 0), compared to the condition where random apical inputs are masked. Masking the context-relevant apical inputs specifically led to a steeper degradation of accuracy as the number of masked inputs increases.}
        \label{fig:mask_analysis}
    \end{figure}
\end{appendixbox}

\end{document}